# Alternative Standard Frequencies for Interstellar Communication

*C Sivaram*

Indian Institute of Astrophysics

Bangalore - 560 034, India

*Kenath Arun*[*]

Christ Junior College

Bangalore - 560 029, India

*Kiren O V*

St. Josephs Indian Composite PU College

Bangalore - 560 001, India

**Abstract:** The 21 cm hydrogen line is considered a favourable frequency by the SETI programme in their search for signals from potential extra-terrestrial civilizations. The Pioneer plaque, attached to the Pioneer 10 and Pioneer 11 spacecraft, portrays the hyperfine transition of neutral hydrogen and used the wavelength as a standard scale of measurement. Although this line would be universally recognized and is a suitable wavelength to look for radio signals from extraterrestrials, the presence of ubiquitous radiation from galactic hydrogen could make searches a little difficult. In this paper we suggest several alternate standard frequencies which is free of interference from atomic or molecular sources and is independent of any bias.

---

[*] Corresponding author:
e-mail: kenath.arun@cjc.christcollege.edu
Telephone: +91-80-4012 9292; Fax: +91-80- 4012 9222

As is well known, more than five decades ago it was suggested that the 21 cm wavelength of the hyperfine spin-flip transition of H atoms (corresponding to a frequency of 1420 MHz) be used for radio communication (both for transmission and receiving of signals) with possible technologically advanced extraterrestrial (ET) civilizations (Cocconi and Morrison, 1959). Hydrogen is the most abundant element (comprising ~ 75%) and the wavelength corresponding to the well-known transition would be recognized by all ET's with advanced astronomy knowledge. In fact emission of this line by cold H atoms in our galaxy has enabled mapping of the Hydrogen clouds in interstellar space, in spiral arms (of our galaxy) enabling complex contour mapping in the radio spectrum. However as the transition amplitude is very weak, the total radiation falling on the earth (for instance) at this wavelength is hardly one watt (from the entire galaxy). The background flux is less than $10^{-21} W/cm^2/str$. (Burke and Graham-Smith, 2002)

Although this line would be universally recognized everywhere (at all epochs) and is a suitable wavelength to look for radio signals from ET's, the presence of ubiquitous radiation from galactic hydrogen (with the Doppler effect due to sources moving randomly) could make searches a little difficult as corrections for the background, motions, etc. have to be made. Indeed the results of initial searches like that of Drake, and others as well as other SETI efforts (at this wavelength) were negative (Heidmann and Dunlop, 1995). It was suggested that the hydroxyl (OH) line at 1612 MHz (18cm) be also used, so that the band covering H and OH, came to be known as 'waterhole' (of the radio spectrum) at which different ET's would meet to communicate. (Ross, 2009; Sivaram and Sastry, 2004)

But at these wavelengths, there could be interference from atomic or molecular species (not to think of Motorola's Iridium satellite which was earlier in the hydroxyl band) jamming possible ET signals. One frequency which could be unique and free of such problems, which has been suggested is the 2556.8MHz, corresponding to:

$$f_1 = \alpha^4 \left( \frac{c}{2\pi r_B} \right) = 2556.8 \ MHz \qquad \ldots (1)$$

where $r_B = \hbar^2/m_e c^2$ is Bohr radius, $\hbar$ is the Planck's constant, $m_e$ is the electron mass, $c$ is speed of light, $\alpha = \frac{e^2}{\hbar c} = \frac{1}{137}$ is the universal number known to all physicists in universe (irrespective of units).

It is interesting that $2\pi r_B/c$, the time taken by light to go around Bohr hydrogen atom turns out to be an attosecond ($10^{-18}$ s). Thus $c/2\pi r_B = 10^{18} Hz$, would be known to all physicists, but the frequency corresponds to X-rays (about 3 angstrom). This would be difficult to use for interstellar communication, because of the cosmic X-ray background and many other sources emitting around this frequency. For similar reasons the electron – positron annihilation gamma ray line of frequency $m_e c^2/\hbar = 10^{21} Hz$, is also not suitable.

The classical electron radius $r_e = e^2/m_e c^2 = 3 \times 10^{-13} cm$, also corresponds the gamma ray wavelength of $10^{23} Hz$, corresponding time interval being ten yoctoseconds. This is again too high a frequency (difficult to generate). Thus in the radio part of spectrum (traditionally and still the most favoured) the only suitable radio frequency (arrived at from only fundamental constants) is:

$$f = \alpha^4 \left( \frac{c}{2\pi r_B} \right) = \alpha^6 \left( \frac{c}{2\pi r_e} \right) = 2556.8 \ MHz \approx 11.8 \ cm \qquad \ldots (2)$$

No other power of $\alpha$ is invoked. There is nothing else radiating in this frequency. No known molecular or atomic lines at this frequency. No recombination lines near this frequency. So no interference from atomic or molecular species, and this frequency is independent of any assumptions about chemistry, i.e. bias free. Moreover there is minimal galactic background noise at this frequency. Thus, for transmission, power requirements are less stringent.

Can other combinations of fundamental constants lead to possible radio frequencies? Since astronomers are strongly convinced about the dominance of dark energy in the universe, it is conceivable that ET astronomers should also have arrived at the same conclusion. Observations suggest that dark energy could very well be the cosmological constant $\Lambda$, with value of $\Lambda = 10^{-56} cm^{-2}$ (Arun et al, 2017). The combination of Planck length, $l_{Pl} = (\hbar G/c^3)^{1/2} = 1.6 \times 10^{-33} cm$ (where, $\hbar$ is the Planck's constant, $G$ is the gravitational constant and $c$ is the speed of light), and $\Lambda$ gives a length (wavelength): (Sivaram, 1986a; 1986b)

$$l = \left(\frac{\hbar G}{c^3 \Lambda}\right)^{1/4} = 3 \times 10^{-3} cm \qquad ... (3)$$

corresponding to a frequency of 10 THz. This is not suitable possibly as lot of foreground noise from interstellar grains (with temperatures of 50 K) could obscure the signal. Although, THz generation is picking up (we have TASERS) for various applications on earth. Again the combination of beta decay length, $l_w = (G_F/\hbar c)^{1/2} \approx 7 \times 10^{-17} cm$ (where $G_F$ is the Fermi weak interaction constant, and is a universal constant and would be known to all advanced ET physicists), and Planck length lead to:

$$\frac{l_w^2}{l_{Pl}} \approx 3 cm \qquad ... (4)$$

This gives a frequency of 10 GHz, which is suitable, falling within the atmospheric window. Also combination of $l_w$ and $\Lambda$ gives:

$$\frac{c}{(G_F/\hbar c \Lambda)^{1/4}} \approx 50 \ kHz \qquad \ldots (5)$$

which is too low a frequency. (Sivaram et al, 2017)

Combination of other terms (i.e. nuclear radius, which is again a universal parameter) and Λ gives rise to a frequency of ~1 *kHz*. The latter two frequencies are suitable for a radio telescope on the far side of moon and could well be the frequencies at which ET's may choose to communicate (Ironically 1 kHz, is also the frequency at gravitational waves associated with stellar collapse (Hawking and Israel, 1979), as also near to frequencies emitted by the brain), though the physics is different.